\documentclass[12pt,a4paper]{article}
\usepackage{jheppub}
\usepackage[T1]{fontenc} 
\usepackage{epsfig} 
\usepackage{amscd}
\usepackage{verbatim}
\usepackage{latexsym}
\usepackage{amsfonts,amsthm,amsmath,mathtools}

\notoc 

\title{4d/3d reduction of s-confining theories:\\
the role of the "exotic" D instantons}
\author[a]{Antonio Amariti,}

\affiliation[a]{Laboratoire de Physique Th\'eorique de l'\'Ecole Normale Sup\'erieure \\
24 Rue Lhomond, Paris 75005, France}

\emailAdd{amariti@phys.lpt.ens}

\abstract{
In the reduction of 4d dualities to 3d there are 
non-perturbative effects arising from monopoles acting as instantons.
This mechanism has been reproduced in string theory by engineering 
the theories in a IIA  brane setup.
Nevertheless there are limiting cases of the 4d dualities
where the dual theories are actually confined phases of the UV gauge theories. 
In these cases the monopoles are absent and the mechanism of reduction 
of the 4d duality has to be  modified.
In this paper we investigate such modification in the brane setup.
The main observation behind our analysis is that in the 4d case the superpotential
of the confined theories can been obtained also as the "exotic" contribution
of a D0 brane, a stringy instanton.
When  considering these configurations 
we reproduce  the field theory results  in the brane setup.
We study both the unitary and the symplectic case. 
As a further check we study the reduction of the 4d
superconformal index to the 3d partition function for these theories.
  \vspace{6cm}
}

\begin{document}

\maketitle \newpage
\tableofcontents

\section{Introduction}

In this paper we study the dimensional reduction
of $\mathcal{N}=1$ 4d s-confining theories
to 3d in the brane setup.
We show that a key role 
is played by the exotic contribution of stringy instantons.

A general procedure for reducing 4d dualities to 3d has been furnished
in \cite{Aharony:2013dha,Aharony:2013kma}  (see also \cite{Niarchos:2012ah} for an earlier discussion). It is based on the observation that a straight compactification
of dual theories on a circle generally spoils the 4d duality. This is because when reducing the theories the scale of validity of the 3d duality is lower than the effective scale at which the new theories have to be considered.
Alternatively one can think that in this process extra symmetries, anomalous in 4d, are generated in 3d. These symmetries lead the 4d dual theories to different fixed points  in 3d.

This problem can be avoided if the finite size effects of the circle are considered.
In this way there is an effective 3d duality at the scale set by the dimension of the circle.
This is the scale of the KK monopoles, that act as instantons in this setup and generate 
non perturbative superpotentials. These superpotential break the extra symmetries
discussed above.
The theories on the circle can be considered as effective 3d dualities. More conventional dual pairs are obtained by a real mass flow. 

This procedure is general and can be applied to any 4d duality between 
gauge theories. It has been shown  \cite{Amariti:2015yea,Amariti:2015mva} that if the theories have a IIA brane realization the mechanism can be reproduced  in string theory.  This is based on T-duality and the effects of the monopoles are reproduced by
the action of D1 branes.

There is another possibile IR behavior of UV free theories in 4d: confinement. 
In this case the low energy dynamics is not described by a dual gauge theory but in terms of  mesons and baryons. 
In this case the reduction to 3d is more complicated and the prescription
of \ \cite{Aharony:2013dha,Aharony:2013kma} requires some modification. 
Before describing the reduction in the confining phase we discuss some 
4d aspects of these theories that will be useful in the following. 

In supersymmetry some confining theories correspond to limiting cases of  electric/magnetic Seiberg dualities \cite{Seiberg:1994pq}.
The simplest example is $\mathcal{N}=1$ SQCD with $N$ colors and  $N+1$ flavors.
This theory is s-confining in the IR \cite{Seiberg:1994bz}, and in this regime 
it is described by the mesonic and the baryonic degrees of freedom, interacting through a superpotential.
This superpotential corresponds to the classical constraint on the moduli space.
Equivalently this low energy description has been obtained by adding a massive flavor in the
gauge theory and by studying the large mass regime \cite{Seiberg:1994pq}.
In this case the theory has originally $N+2$ flavors and the strongly coupled phase can be described by 
a dual IR free $SU(2)$ gauge theory. Integrating out the massive flavor in the electric theory corresponds to 
a total higgsing of the $SU(2)$ dual gauge group. 
In this case the superpotential is reproduced by a scale matching relation on the instantonic 
contribution of the totally broken $SU(2)$.

Gauge instantons have a counterpart in string theory: they are related to 
Dp branes placed on stacks of D(p+4) branes (see \cite{Tong:2005un,Blumenhagen:2009qh} 
and references therein for review). The stack of D(p+4) branes corresponds to the non-abelian 
gauge theory and the Dp branes reproduce the effect of the gauge instantons.

The gauge instanton effect discussed above can be captured in a different way in string theory,
without requiring to UV complete the confining phase to an $SU(2)$ gauge theory. 
This effect can be observed by a IIA brane engineering, by considering 
a single D4 brane extended between two non parallel NS branes, with in addition some D6 flavor branes. 
In 4d the abelian gauge factor associated to this D4 brane decouples in the IR.
Unexpectedly, also in this case, the D instanton, an euclidean D0 brane,
reproduces the superpotential effect of the 
gauge instanton of the broken $SU(2)$
\cite{GarciaEtxebarria:2007zv,Krefl:2008gs,Amariti:2008xu}.
For this reason this D instanton has been called
"exotic" in the literature. 
Observe that the size of the instantonic correction 
is different in the two regimes, the stringy and the field theory one, 
signaling that the two descriptions are accurate at
different scales \cite{Krefl:2008gs}.
\\
\\
In this paper we study the fate of this type of  stringy instantons when the s-confining
gauge theories are compactified on a circle and reduced to 3d.
We perform the reduction separately in the confining and in the confined phase\footnote{Observe that in 4d, 
in  the limiting case of $SU(N)$ Seiberg duality, the
electric theory becomes a confining phase while the magnetic theory is
identified with the confined phase. 
In the 3d case, where the theories are conformal, we have a duality. For this reason, with a slight 
abuse of notation, in the 3d case, we refer to the two dual phases as to the electric and the magnetic
theory.} 
both in the field theory regime and in the string theory regime.
In the field theory regime we follow the procedure of \cite{Aharony:2013dha,Aharony:2013kma} 
for reducing the confining phase.
 In the confined regime the gauge theory is absent. In this case there is a
 different prescription  \cite{Csaki:2014cwa,Amariti:2015kha} stating that the effective confined theory on the circle has the same field content and superpotential of the 4d parent.
 In the string theory regime we follow the arguments of \cite{Amariti:2015yea,Amariti:2015mva} for reducing the confining gauge 
theory on the circle.
In the confined case we observe that the extra contribution 
is captured by the T-dual version of the D0 stringy instanton. This corresponds to a D1 euclidean brane, or a D string, 
\emph{i.e.} a monopole acting as an instanton in 3d. 
Thanks to this observation we reproduce the results obtined in field theory
with the prescription of \cite{Csaki:2014cwa,Amariti:2015kha}.

We study the reduction of 4d confining theories both with unitary and symplectic 
gauge groups.
Furthermore we study this mechanism in terms of the reduction of the 4d
superconformal index to the 3d partition function.

In section \ref{sec:4ds} we review the 4d s-confining $SU(N)$ SQCD and the relation with the stringy instanton.
In section \ref{sec:redSU} we study the reduction of this theory to 3d in field theory and in string theory. 
We study the role of the stringy instanton in the brane engineering of this theory.
In the 3d limit, by gauging the baryonic symmetry, 
 we arrive at the $U(N)$ case with $N$ flavors, 
and reproduce the limiting case of Aharony duality
\cite{Aharony:1997gp}. 
We conclude this section with the reduction of the superconformal index to the 
partition function. It confirms the validity of the procedure. 
In section \ref{sec:redSP} we repeat the analysis for the symplectic case.
In section \ref{sec:conc} we conclude.

\section{s-confinement and exotic instantons}
\label{sec:4ds}
In this section we review the exotic effects of stringy instantons in 4d $\mathcal{N}=1$ supersymmetric gauge theories.
This instanton configuration has been extensively studied in cascading  quiver gauge theories
\footnote{Simliar ideas have been discussed in the context of matrix model
\cite{Aganagic:2003xq}. The relation with the stringy instanton has been 
shown in \cite{GarciaEtxebarria:2008iw}.}, 
associated to the CY singularity probed by a stack of D3 brane in AdS/CFT 
\cite{GarciaEtxebarria:2007zv,Florea:2006si,Argurio:2007vqa,Aharony:2007pr,
Petersson:2007sc,Kachru:2008wt,
Argurio:2008jm,Ferretti:2009tz}.
The RG cascading quivers are obtained by the addition of fractional D3 branes. 
There are cases where some of the nodes have rank $N=1$, i.e. a single D3 is left at the end of the duality cascade 
on such nodes.
With an abuse of notation we denote these nodes as $SU(1)$ nodes, having in mind the decoupling of the $U(1)$ factors in the IR.
Even if there are no gauge dynamical degrees of freedom from the singularities associated to these nodes, the latter play a role in the
dynamics if euclidean rigid D(-1) instantons are wrapped on them.
The instantons generate a non perturbative dynamics that modifies the effective superpotential.
This is obtained by considering the bosonic and fermionic zero modes 
in the ADHM construction and the relative action and constraints.
Many of the zero modes are lifted, except for two fermionic  modes that correspond to two 
fields, $\alpha$ and $\beta$, also called Ganor strings \cite{Ganor:1996pe},
 connecting the $SU(1)$ node to the rest of 
the quiver. These modes are lifted by an interaction $\alpha M \beta$,  giving origin to the superpotential 
\begin{equation}
\label{eq:W-stringy-unitary}
W = \int d \alpha \, d\beta \, e^{\alpha_a M_{ab} \beta_b} \simeq \det M
\end{equation}
Observe that the fermionic zero modes and the generalized meson $M$ have an index structure inherited from the quiver.

This is reminiscent of the mesonic superpotential appearing in the confined phase
of $SU(N)$ SQCD with $N+1$ flavors
\cite{GarciaEtxebarria:2007zv,Krefl:2008gs,Amariti:2008xu}.
As discussed in the introduction this theory confines in the IR and it can be obtained as 
a limiting case of Seiberg duality. The low energy theory in this case has superpotential of the form
\begin{equation}
\label{eq:W-IR-SU(N)-SQCD}
W = b M \tilde b + \det M
\end{equation}
The second term in (\ref{eq:W-IR-SU(N)-SQCD}) has been interpreted as a gauge instanton in field theory,
by UV completing the s-confining phase to $SU(N)$ SQCD with $N+2$ flavors. 
In this case there is a Seiberg dual description in terms of an $SU(2)$ gauge theory with 
$N+2$ dual flavors. If a mass term $W = m Q_{N+2}^{\alpha} \tilde Q_{\alpha}^{N+2}$
is added the IR theory has $N+1$ light flavors. In the magnetic theory the
mass term enforces the total higgsing of the $SU(2)$ gauge group, forced by the $F$ term of
the meson $M_{N+2}^{N+2}$.
In the higgsed phase the dual quarks are identified with the baryons of the electric theory
with $N+1$ flavors.
The gauge instanton associated to the broken of $SU(2)$
generates a contribution proportional to $\det M$ in the superpotential.
This construction reproduces the superpotential (\ref{eq:W-IR-SU(N)-SQCD}).

As discussed above this contribution can be obtained also from an instantonic calculation
by engineering the gauge theory in an Hanany-Witten (HW) \cite{Hanany:1996ie}
setup.
The electric SQCD theory is represented as the low energy limit of a stack of $N$ 
D4 branes in type IIA string theory.
Here we review the construction of this theory.
\begin{figure}
\begin{center}
\includegraphics[width=10cm]{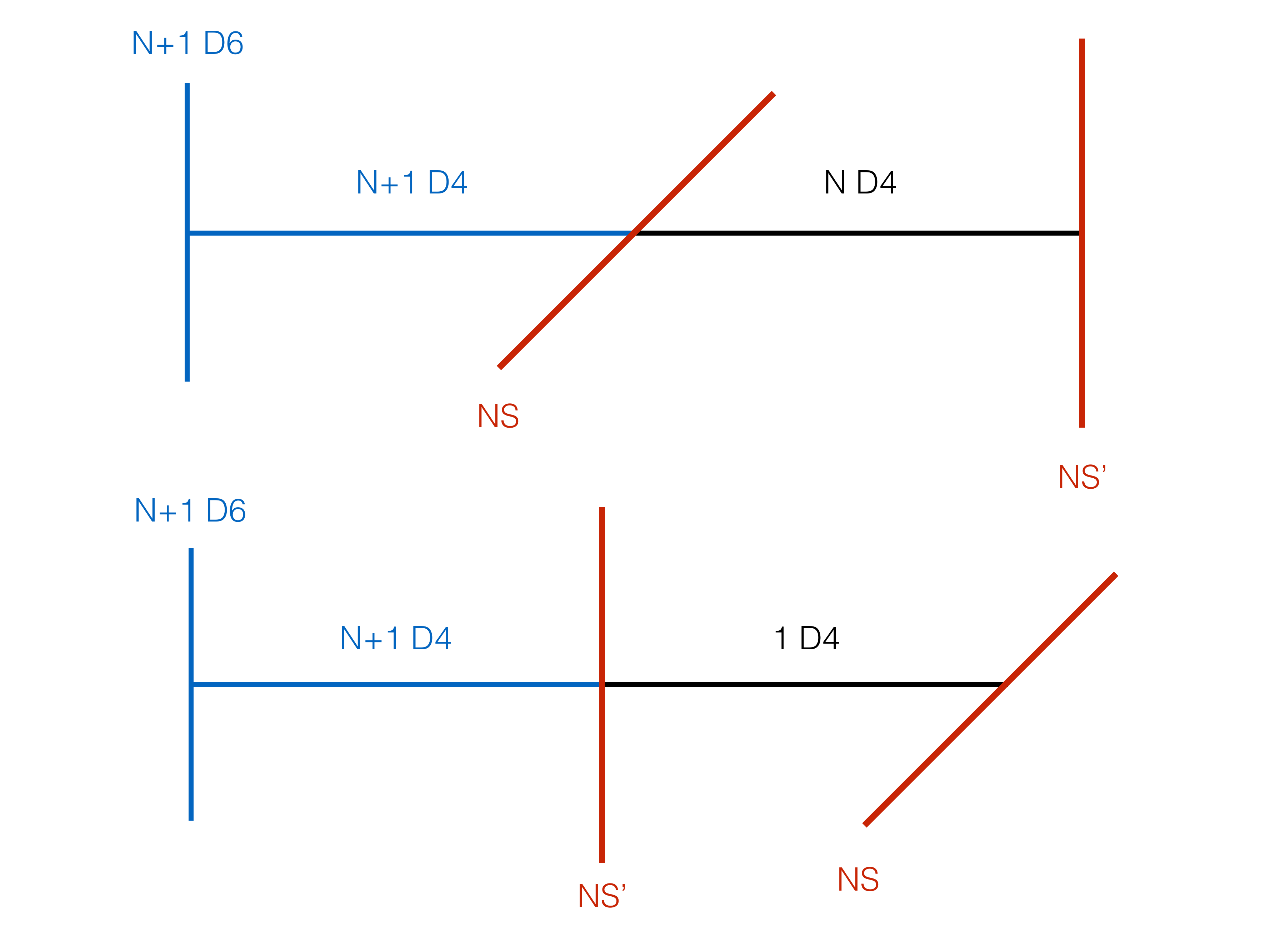}
\caption{The top figure represents the IIA brane setup describing $SU(N)$ SQCD 
with $N+1$ flavors, the bottom figure represents the theory after an HW transition.}
\label{fig:SU(N)-SQCD anD branes}
\end{center}
\end{figure}
The 4d gauge theory is described by a stack of $N$ D4
branes suspended between an NS and an NS' brane. When $N+1$ D6 are considered this setup
describes  $SU(N)$ SQCD with $N+1$ flavors.
The branes are extended in the ten dimensional space-time as  follows:
\begin{center}
\begin{tabular}{l||cccccccccc}
&0& 1&2&3&4&5&6&7&8&9 \\
\hline
D4  &  $\times$ &  $\times$ &  $\times$ &  $\times$ &&&$\times$&&\\
D6  &  $\times$ &  $\times$ &  $\times$ &  $\times$  &&&&$\times$&$\times$ &  $\times$\\
NS     &  $\times$ &  $\times$ &  $\times$ &  $\times$&$\times$ &  $\times$ \\
NS'    &  $\times$ &  $\times$ &  $\times$ &  $\times$ &&&&&$\times$ &  $\times$\\
D0 &&&&&&&$\times$\\
\end{tabular}
\end{center}
The brane setup is shown in Figure \ref{fig:SU(N)-SQCD anD branes}.
The confined theory is obtained by  exchanging the NS and the NS' branes.
A single D4 brane is stretched between the NS and the NS' brane after the transition. This corresponds
to a $U(1)$ gauge theory. In 4d the vector multiplet of an abelian gauge theory decouples in the IR. We are in presence of the $SU(1)$ factor discussed above. 

The light degrees of freedom in this setup are the baryons, oriented
strings connecting the $N+1$ flavor D4 branes
with the
D4 brane stretched between the NS and the NS' brane,
and the meson, open strings with both the endpoints on the $N+1$ flavor D4 branes.
The meson is dynamical because it is associated to the freedom in moving the $N+1$ 
flavor D4 branes in the directions $(8,9)$. This motion corresponds to an effective mass for the baryons and it 
can be represented as a superpotential term 
\begin{equation}
\label{WMbb}
W=M b \tilde b
\end{equation}
In other words the dynamical mass term of the baryon corresponds to the vev of the meson $M$
in the $(8,9)$ directions.
The description of the confined theory in terms of D branes requires also a term proportional to $\det M$ in (\ref{WMbb}). 
This term is obtained from the action of the exotic instanton, \emph{i.e.} an euclidean D0 brane sit on the D4 brane stretched between the NS and the NS' branes. 

\section{Reduction to 3d of $SU(N_c)$ SQCD with $N_c+1$ flavors}
\label{sec:redSU}
In this section we reduce the 4d theories discussed above to 3d, in the field
theory and in the string theory regime.
We also study the reduction of the 4d superconformal 
index to the 3d partition function.

\subsection*{Dimensional reduction in the field theory regime}

When reducing a 4d duality to 3d a straight compactification
may be too naive, and the 3d pairs obtained 
are not necessarily dual.
Indeed 4d anomalous abelian symmetries may arise in 3d
and mix with the current $J_\mu^R$. This mixing can 
induce different RG flows in the two phases, spoiling the original
4d duality \cite{Aharony:2013dha}.

There is a general procedure to reduce a 4d dual pair to a 3d one.
In consists of considering the 4d theories on $\mathbb{R}^3 \times S^1$
as effective 3d theories. The finire size effects from the circle are encoded into
an effective superpotential \cite{Aharony:2013dha}.
This effective interaction prevents the generation of 4d anomalous symmetries in 3d.
This procedure generates 3d dual effective theories.
More conventional dual pairs are recovered by shrinking the circle $S^1$: this
limit is not always possible. In some cases, like the $SU(N)$ and $SP(N)$ 
theories considered here, it requires a further  real mass deformation, 
leading to an RG flow.

This general procedure is valid when both the phases of the duality correspond to a gauge theory.
In these cases the extra non perturbative effects can be thought as the fractionalization
of a 4d instanton into 3d monopoles \cite{Brodie:1998bv}. These monopoles, acting as instantons in 3d,
are of two types. There are BPS monopoles, that survive the compactification and KK monopoles,
that encode the information of circle at finite radius 
\cite{Hanany:1996ie,deBoer:1997ka,Davies:1999uw,Davies:2000nw}.

In the limiting cases considered here, 
where instead of a  dual theory the IR physics is described by a confined phase, 
there are no instanton contributions in the low energy theory, because the gauge 
group vanishes.
Anyway there is a contribution to the superpotential of non-perturbative origin. This  
effect corresponds to the instanton of the totally broken $SU(2)$ as discussed above. 
One can study the $SU(2)$ dual gauge theory on  $\mathbb{R}^3 \times S^1$,
consider the effect of the KK monopole and 
completely break the dual gauge symmetry as done in the 4d case.
A different strategy has been proposed in \cite{Csaki:2014cwa}: when reducing a confined phase on the circle 
the effective theory is formally identical to the 4d parent.  We will adopt this strategy in 
the rest of the discussion.

First we reduce the electric gauge theory, $SU(N)$ SQCD with $N+1$ flavors,
on $\mathbb{R}^3 \times S^1$. 
Here there is a KK monopole contribution, through the effective superpotential
\begin{equation}
\label{eq:eta}
W=\eta Y \quad \text{where} \quad
Y=e^{(\sigma_1-\sigma_{N_c})/g_3^2+i(\varphi_1-\varphi_{N})}.
\end{equation}
We will refer to this superpotential as the $\eta Y$ superpotential 
in the rest of the paper.
 The fields $\sigma$  are the real scalars in the vector multiplet and the fields
$\varphi$ correspond to the dual photons, in the Coulomb branch. These two fields organize in a chiral
multiplet $\Sigma=\sigma/g_3^2+i \varphi$
that parameterizes the directions of the Coulomb branch. The operator $Y= e^{\Sigma}$ corresponds to a monopole operator in the
high energy description.
The 4d gauge coupling $g_4$ reduces to the 3d one by the 
relation $g_4^2 = 2 \pi r  g_3^2$. The extra superpotential (\ref{eq:eta})
is associated to the holomorphic scale of the 4d theory by 
$\Lambda^b = \eta = e^{\frac{4 \pi}{r g_3^2}}$.
The confined case on $\mathbb{R}^3 \times S^1$ corresponds to 
the  set of mesons and baryons discussed above. 
The superpotential of this theory is again  (\ref{WMbb}).
 
The pure 3d duality is obtained from the duality on the circle 
by perturbing the electric theory with a real mass deformation.
We assign one large real mass to a pair of fundamental and anti-fundamental
and reduce the flavor from $N+1$ to $N$.
This is done by weakly gauging  a combination of generators of the baryonic symmetry and of the non abelian symmetry as discussed in the appendix of \cite{Csaki:2014cwa}.
We choose a combination that assigns the opposite sign to the real masses of the two fields.
In the large mass limit the electric theory becomes $SU(N)$ SQCD with $N$
flavors. The masses have opposite sign and the flow does not generate any CS term.
The real mass deformations induces real masses also in some components of the
mesons and of the baryons. These masses are 
assigned consistently with the global symmetry structure.
In the dual theory, if we split the fields as 
\begin{equation}
M = 
\left(
\begin{array}{cc}
M_{i}^{i} & M_{i}^{N+1}\\
M_{N+1}^{i}&M_{N+1}^{N+1}
\end{array}
\right)
\quad 
B = \left(
\begin{array}{cc}
B_{i} &B_{N+1}
\end{array}
\right)
\quad
\tilde B =
\left( 
\begin{array}{l}
\tilde B^{i}\\
\tilde B^{N+1}
\end{array}
\right)
\end{equation}
the massless components are
$M_{i}^{i}$,$M_{N+1}^{N+1}$, $B_{N+1}$ and $\tilde B^{N+1}$.
The superpotential for the massless fields is
\begin{equation}
W =  M_{N+1}^{N+1} (B_{N+1} \tilde B^{N+1} +  \det M_{i}^{i})
\end{equation}
The singlet $M_{N+1}^{N+1}$ has the same global charges of the 
electric monopole $Y$ defined in (\ref{eq:eta}).
We identify $M_{N+1}^{N+1}$ with $Y$ and obtain the 3d duality 
corresponding to the limiting case of $SU(N)$ Aharony duality \cite{Aharony:1997gp}.
 
We can also gauge the $U(1)_B$  baryonic symmetry. The electric theory in this case becomes 
$U(N)$ SQCD with $N$ flavors.  In the dual phase we have a $U(1)$ gauge theory with
one charged fundamental and one charged anti-fundamental.
This theory is mirror dual to the $\mathcal{XYZ}$ model \cite{Aharony:1997bx}.
Here we associate the field $\mathcal{X}$ to the gauge invariant combination 
$\tilde B^{N+1} B_{N+1}$, while the other two chiral fields 
can be denoted as $\mathcal{Y}=v_+$ and $\mathcal{Z}=v_-$. The superpotential of the mirror 
dual theory becomes
\begin{equation}
W = Y \mathcal{X} + Y \det M_{i}^{i} + v_+ v_- \mathcal{X} 
\end{equation}
By integrating out the massive fields we obtain the relation $Y=v_+ v_-$ .
Eventually the superpotential
of the dual theory is
\begin{equation}
\label{Wfin}
W = v_+ v_- \det M_{i}^i
\end{equation}
and it corresponds to the superpotential of the
limiting case of $U(N)$ Aharony duality \cite{Aharony:1997gp}.
\subsection*{Brane interpretation}

Here we provide a brane interpretation of the reduction discussed above.
A general procedure for reducing 4d dualities engineered in type IIA
string theory to 3d dualities in IIB setups has been developed in \cite{Amariti:2015yea}
for unitary theories with fundamental flavor,  and extended in \cite{Amariti:2015mva}
to more general gauge and field content.
The procedure is based on compactification and T-duality.
The KK monopole effects are captured by D1-branes
or, by S-duality, by F-strings.
Reducing to pure 3d pairs requires a double scaling limit on the
radius and the real masses, associated to the position of some
flavor brane on the circle.

Here we consider the 4d brane setup of Figure \ref{fig:SU(N)-SQCD anD branes},
and compactify the theory on $x_3$. 
By T-duality along this direction the IIA system becomes a IIB system and describes the
effective 4d field theory on $R^3 \times S_r^1$, where $r$ is the radius of the circle.
The NS and NS' branes are left invariant by T-duality  while the D4 and the D6 branes
are become D3 and D5 branes respectively. The D0 branes become D1 branes.
At large T-dual radius $\alpha'/r$ this brane setup describes an effective 3d theory.

When considering a  3d duality at brane level we must also gauge the $U(1)_B$ symmetry \cite{Amariti:2015yea}.
At brane level we associate this symmetry to the relative position of the center of mass
of the stack of the gauge D3 branes with respect to the position of the center of mass of the flavor D3/D5 branes.
By fixing the position of the center of mass of one stack and allowing the motion of the other
one can consider the $U(1)_B$ symmetry as gauged or not.

In the $U(N)$ theory on the circle the effect of the monopoles  
is encoded in the D1 branes
stretched between the D3 and the NS branes  along the directions $x_3$ and $x_6$. Equivalently 
it is associated to spectrum of the S-dual F-strings.
In the 3d decompactified case these effects, corresponding to the repulsive interactions between the parallel
D3 branes, are associated to the BPS monopole superpotential, $W \simeq \sum Y_i^{-1}$. In the compact case there is an additional contribution, corresponding to the KK monopole contribution 
\cite{deBoer:1997ka,Davies:1999uw}.

The analysis of the reduction of the confining phase differs from the one of \cite{Amariti:2015yea}.
Indeed in this case the $\eta Y$ superpotential cannot arise, because of the absence of the dual gauge group.
In the brane picture there is still a D3 brane and we can consider the effect of a D1
brane  wrapping the $x_3$ direction.
This is the effect of the 4d stringy instanton once the theory is reduced on the circle.
In 4d it gave origin to the extra term  $\det M$ in the superpotential.
Here, in the 3d description, this effect is captured by the T-dual D1 brane.

We can further flow to a pure 3d duality by a real mass flow. 
In the electric case we  move a D5 brane at $x_3 = \pi r$ on the circle.
By fixing the D3 branes at the position $x_3 = 0$ on the T-dual circle
we locate a D5 at the position $x_3 = \pi \alpha'/r$, defined as the mirror point in \cite{Amariti:2015mva}.

By considering the limit $r \rightarrow 0$ the sector at this mirror point can be further decoupled
and we are left with 3d $U(N)$ SQCD with $N$ flavors.
In the dual theory the motion of a D5 at the mirror point generates a D3 brane by the HW effect.
In this case there are no D3 branes left in the gauge sector at $x_3=0$. 
The baryons at $x_3=0$ are massive and there is an $N \times N$ meson $M$ left. 
At small $r$ an S-duality produces an $\mathcal{XYZ}$-like model 
at the mirror point. This is not the usual $\mathcal{XYZ}$ model because there are $N$ D3
stretched between a D5 parallel to an NS' brane. This signals the presence of 
a mass term for one of the singlets, the freedom to move the D3 branes between the D5 and the parallel NS'.
The superpotential is $W=\mathcal{X} Y +  \mathcal{XYZ}$,
where $Y$ parameterizes the motion of the flavor D3 brane placed at the mirror point.

 In the small $r$ limit this field $Y$
interacts with the mesons $M_{11}$ at $x_3=0$ through the superpotential generated by the D1 brane.
This interactions corresponds indeed to moving one of the D5 brane at $x_3=\pi \alpha'/r$.
The original term $\det M$ becomes in this case $Y \det M_{1}^{1}$.
Putting everything together the superpotential becomes 
$W=\mathcal{YZ} \det M_{11}$, that corresponds to (\ref{Wfin}) if the fields $\mathcal{Y}$ and 
$\mathcal{Z}$ are identified with the monopole
and the anti-monopole of the $U(N)$ theory, $v_{\pm}$.
\subsection*{The partition function}

The reduction of 4d $SU(N)$  SQCD with $N+1$ flavors can be studied also at the level of the 
4d superconformal 
\footnote{We keep the usual abuse of notation in this terminology because
the index does not require a superconformal theory
\cite{Festuccia:2011ws}, but just the presence of a conserved R symmetry.
More correctly we should refer to the supersymmetric partition function on $S^3 \times_q S^1$.}
index \cite{Kinney:2005ej,Romelsberger:2005eg}. 
The index reduces to the 3d partition function 
  \cite{Dolan:2011rp,Gadde:2011ia,Imamura:2011uw,Agarwal:2012hs}, 
  computed on a squashed three sphere $S_b^3$
\cite{Hama:2011ea}, preserving an
$U(1)^2$ isometry of $SO(4)$. The subscript $b$ represents the squashing parameter.

The index for the 4d theories has been studied in \cite{Spiridonov:2009za,Spiridonov:2014cxa}. 
After we reduce the 4d index 
we obtain the 3d partition function of $SU(N)$ SQCD  with $N+1$ flavor with the
extra $\eta Y $ superpotential. This partition function can be written as
\begin{eqnarray}
\label{eleS1}
\mathcal{Z}_e(\mu+m_B;\nu-m_B) &=& \int \prod_{i=1}^{N} d\sigma_i 
\prod_{a=1}^{N+1}\Gamma_{h}(\mu_a+\sigma_i+m_B)
\Gamma_{h}(\nu_a-\sigma_i-m_B) \nonumber \\
&\times&
\prod_{1\leq i<j \leq N}  \Gamma_{h}^{-1} (\pm(\sigma_i-\sigma_j))
\delta \Big(\sum_{i=1}^{N} \sigma_i\Big)
\end{eqnarray}
The functions $\Gamma_h$, hyperbolic gamma functions \cite{vandebult},  are the one loop-exact determinants
of the matter and vector multiplets 
computed from localization in \cite{Hama:2011ea,Jafferis:2010un,Hama:2010av}.
The parameters $\mu_a$ and $\nu_a$ correspond to the real masses associated to the 
$SU(N+1)_l \times SU(N+1)_r$ flavor symmetry. These masses have an imaginary part
corresponding to the $R$-symmetry, i.e. $\mu_a = m_a + \omega \Delta$ and 
$\nu_a = \tilde m_a + \omega \Delta$, where 
$m_a$ and $\tilde m_a$ are real mass parameters satisfying $\sum m_a=\sum \tilde m_a = 0$, 
$\Delta$ is the $R$-charge and $\omega \equiv i( b+1/b)$. 
The parameter $m_B$ is associated to the
real mass for the baryonic $U(1)_B$ symmetry. The real coordinate $\sigma_i$  
corresponds to the scalar in the vector multiplet, and the constraint $\sum \sigma_i=0$ is
enforced by the $\delta$-function.
The real scalar parameterizes the fundamental representation as  $(+ \sigma_i)$,
the anti-fundamental as $(-\sigma_i)$ and the adjoint as $(\sigma_i  - \sigma_j)$. 
We used the definition $\Gamma_h(\pm z) = \Gamma_h(z) \Gamma_h(-z)$.
There is a constraint between the real masses, that corresponds to the condition 
imposed by the presence of the superpotential (\ref{eq:eta}).
This constraint is
\begin{equation}
\label{balancing}
\sum_{a=1}^{N+1} \mu_a + \sum_{a=1}^{N+1} \nu_a = 2\omega
\end{equation}
We can also study the effect of the gauging of the baryonic $U(1)_B$ symmetry on the partition function.
First we introduce a factor $e^{2 \pi i  m_B \Lambda N}$, where $\Lambda$ is an arbitrary real 
parameter.
Then we Fourier transform the $\delta$-function,
we shift $\sigma_i \rightarrow \sigma_i-m_B$ and obtain a new $\delta$-function $\delta (\Lambda -\xi)$.
After performing the integral over $\xi$ we obtain
\begin{equation}
\mathcal{Z}_e(\mu;\nu;\Lambda)   = 
\frac{1}{N} \int e^{2 \pi i \Lambda \sum_i \sigma_i} \prod_{i=1}^{N} d\sigma_i 
\prod_{a=1}^{N+1}\Gamma_{h}(\mu_a+\sigma_i)
\Gamma_{h}(\nu_a-\sigma_i) \!\!\!\!\!
\prod_{1\leq i<j \leq N} \!\!\!\!\! \Gamma_{h}^{-1} (\pm(\sigma_i-\sigma_j))
\end{equation}
When reduced on the circle
the partition function of the 4d confined phase
corresponds to the product of the contributions of the mesons and of the baryons.
The 3d partition function of the dual effective theory is
\begin{equation}
\label{magS1}
\mathcal{Z}_m=\mathcal{Z}_M \mathcal{Z}_b \mathcal{Z}_{\tilde b} = \prod_{a,b}^{N+1} \Gamma_{h} (\mu_a+\nu_b)
\prod_{a=1}^{N+1}  
\Gamma(\omega+N m_B-\mu_a) \Gamma_h(\omega-N m_B-\nu_a)
\end{equation}
The partition function in (\ref{eleS1})
coincides with the one in (\ref{magS1})
if the parameters satisfy (\ref{balancing}). 
As done in the electric case we can further add the extra factor  $e^{2 \pi  i m_B \Lambda N}$ 
and gauge the $U(1)_B$ symmetry.
We obtain
\begin{equation}
\mathcal{Z}_m = \prod_{a,b}^{N+1} \Gamma_{h} (\mu_a+\nu_b) 
\int d m_B e^{2 \pi i N m_B \Lambda}\prod_{a=1}^{N+1}  
\Gamma(\omega+N m_B-\mu_a) \Gamma_h(\omega-N m_B-\nu_a)
\end{equation}

The decompactification limit requires a real mass flow
in the field theory analysis.
This real mass flow is reproduced if we consider the assignation
\begin{eqnarray}
\mu_a =
\left\{ 
\begin{array}{ll}
m_a +m_A & a=1,\dots, N\\
m - m_A N + \omega &
\end{array}\right.
\quad\quad
\nu_a =
\left\{ 
\begin{array}{ll}
~~\tilde m_a +m_A & a=1,\dots, N\\
-m - m_A N +\omega&
\end{array}\right.
\nonumber \\
\end{eqnarray}
with the constraint   $\sum m_a=\sum \tilde m_a=0$.
The flow is reproduced by the limit $m\rightarrow \infty$ on the partition function.
On the hyperbolic gamma functions this limit is obtained from the formula
\cite{vandebult}
\begin{equation}
\label{kargeZmass}
\lim_{x \rightarrow \infty} \Gamma_h(x) = e^{i \pi \text{sign}(x) (x-\omega)^2}
\end{equation}
We study this limit on the partition function on both sides of the duality.
The partition function of the electric theory becomes (we omit the large $m$ dependence in 
the following because we checked that it coincides to the one computed in the dual frame)
\begin{eqnarray}
\label{Zelefin}
\mathcal{Z}_e = \!\!\! \int \frac{e^{2 \pi i \Lambda \text{Tr} \sigma} }{N} 
\prod_{i=1}^{N} d\sigma_i 
\prod_{a=1}^{N}
\Gamma_{h}(m_a+m_A+\sigma_i)
\Gamma_{h}(\tilde m_a+m_A -\sigma_i) 
\!\!\!\!\!
\prod_{1\leq i<j \leq N} \!\!\!\!\! \Gamma_{h}^{-1} (\pm(\sigma_i-\sigma_j))
\nonumber \\
\end{eqnarray}
This formula corresponds to the partition function of the $U(N)$ gauge theory with $N$ 
flavors.
In the magnetic case we first rescale $m_B$ as $m_B/N$ and then assign the real masses.
 In the large $M$ limit we have
\begin{equation}
\mathcal{Z}_m = \Gamma_h(2 \omega-2 N  m_A )) \prod_{a,b}^{N} \Gamma_{h} (m_a+\tilde m_b+2 m_A) 
\int \frac{d m_B}{N} e^{2 \pi i  m_B \Lambda} 
\Gamma(\pm m_B+N m_A) 
\end{equation}
The last step consists of using the duality between SQED with one flavor and the $\mathcal{XYZ}$ model.
On the partition function this duality is encoded in the relation \cite{Benini:2011mf}
\begin{equation}\int d m_B e^{2 \pi i  m_B \Lambda}
\Gamma(\pm m_B+N m_A) 
=
\Gamma_h(2 N m_A) \Gamma_h(\pm\frac{\Lambda}{2}- N m_A  +\omega  )
\end{equation}
The product $\Gamma_h(2 N m_A) \Gamma_h(2\omega-2 N m_A) $
is compatible with a superpotential mass term and it can be further simplified by the relation $\Gamma_h(x) \Gamma_h(2 \omega-x)=1$.
We obtain
\begin{equation}
\label{Zmagnfin}
\mathcal{Z}_m =
\prod_{a,b}^{N} \Gamma_{h} (m_a+\tilde m_b+2 m_A) 
\Gamma_h\bigg(\pm\frac{\Lambda}{2}- N m_A  +\omega \bigg )
\end{equation}
that is equivalent to (\ref{Zelefin}) and describes the partition function of the dual theory with superpotential (\ref{Wfin}).
Observe the role of $\Lambda$: it is an FI term in the electric theory, 
added when gauging the baryonic symmetry. 
This parameter becomes a real mass parameter
in the magnetic theory. This is expected because the FI parameter corresponds to the real mass parameter
of the  $U(1)_J$ symmetry. This is a topological symmetry
that shifts the dual photon and that arises only in the $U(N)$ case. In the dual theory the topological
symmetry does not disappear, even if the dual gauge theory is trivial, because
there are gauge singlets, the electric monopole operators, 
carrying a non trivial charge under $U(1)_J$.

\section{The symplectic case}
\label{sec:redSP}

Another exotic instanton contribution has been 
obtained for models with symplectic gauge groups.
In this case a stringy instanton contributes to the effective Lagrangian of 
$SP(0)$ theories (as in the $SU(1)$ case here we use a similar abuse of terminology) and corresponds to an $O(1)$ instanton. 

\subsection*{The $O(1)$ instanton in 4d}

Here we consider a 4d $\mathcal{N}=1$ $SP(2N)$ gauge theory
\footnote{We use the convention $SP(2)\simeq SU(2)$.} with $2(N+2)$ flavors.
The theory confines in the IR \cite{Intriligator:1995ne}. 
In the symplectic case there are no baryons and the low energy description
consists of a mesonic operator $M= Q Q $ with superpotential 
\begin{equation}
\label{Pf}
W = \text{Pf} \, M 
\end{equation}
This theory can be thought as the limiting case of Seiberg duality for an $SP(2N)$ gauge group with $2F$ flavors,
where indeed the dual gauge group is $SP(2(F-N-2))$. For this reason we denote the theory as an $SP(0)$ gauge theory.

It has been shown that in quiver gauge theories, if there are 
confining symplectic groups, 
the effective superpotential (\ref{Pf}) is obtained by wrapping a
D instanton on the singularity associated to that node \cite{GarciaEtxebarria:2007zv,Argurio:2007vqa}.
For example, consider a IIA description of an elliptic quiver, with a circular D4 brane 
intersecting NS and NS' branes.
By adding fractional branes the number of D4 branes between two consecutive and
non parallel NS branes is reduced by an HW transition. 
We can add also orientifolds to this geometry, O4 or O6 planes.
The O4 can be put on the circular D4 branes and it switches its  charge  
each time it crosses an NS branes. In this case we have a quiver with alternating SO/SP groups. On the other hand,
if we can consider the action of the O6 planes there can be both real and unitary gauge groups.

Consider a node of an elliptic quiver associated to a segment with $2N_i$ D4 branes, 
stretched between a pair of NS branes.
When an O4$^+$ or an O6$^-$ plane
\footnote{The sign represents the action of the orientifold 
on the NS sector. The charge is associated to the projection of
an $SU(2N_i)$ gauge theory.}
acts on this node the gauge group is projected  
to $SP(2N_i)$.
If $N_i=0$ the group is $SP(0)$. Nevertheless if we consider a D(-1) brane wrapping this node
we have an $O(1)$ instanton. The orientifold projects out the extra fermionic zero
modes in the ADHM construction and the instanton contributes to the superpotential.
There are only two Ganor strings \cite{Ganor:1996pe} connecting this nodes to the other(s) and the instanton contributes to the superpotential 
with a contribution of the form  (\ref{Pf}), where the meson is obtained in terms of the other bi-fundamentals of the quiver.

\begin{figure}
\begin{center}
\includegraphics[width=10cm]{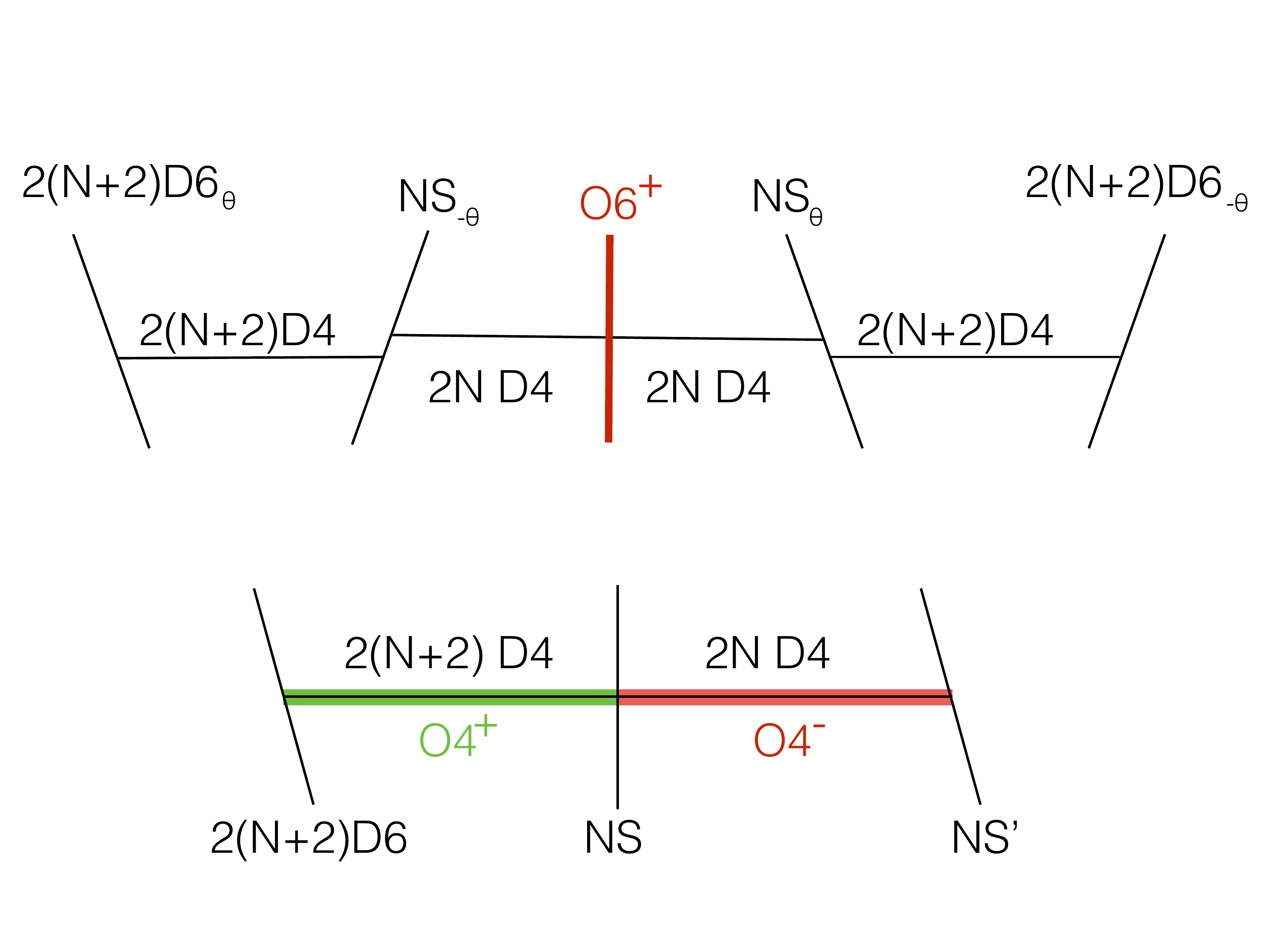}
\caption{In this figure we reproduce the IIA brane setup describing $SP(2N)$ SQCD 
with $2(N+2)$ flavors. We represent both the realizations in terms of the O6 and O4 plane.}
\label{O46}
\end{center}
\end{figure}

This construction can be used also for non quiver theories, for example for the $SP(2N)$ SQCD
with $2(N+2)$ flavors discussed above. There are two possible brane configurations.
In one case we put an O6$^-$ plane orthogonal to the stack of $N+2$ D4 branes.
The orientifold is extended along the direction $(0,1,2,3,4,5)$. The NS and the D6 branes in the setup are in this case rotated along (4,5) and (8,9) to preserve $\mathcal{N}=1$ supersymmetry in four dimensions.
In the second case we consider the setup of section \ref{sec:4ds}
and add on a stack of $2(N+2)$ D4 branes
an O4$^{+}$ planes. It becomes an O4$^-$ plane when the NS brane is crossed. 
The two cases  are summarized in Figure \ref{O46}.
We can add the D instanton
in both cases and perform an HW transition. After the transition the number of D4 branes 
extended between the NS and the NS' brane vanishes
because of charge conservation
\footnote{This charge is the linking number defined in \cite{Hanany:1996ie}. 
In this case the cancellation occurs because the orientifold modifies this charge.}.
Even in absence of D4 branes the D instanton contributes to the superpotential.
The exotic contribution is  
\begin{equation}
\label{WSP}
W = 
\int d \alpha \,
e^{\alpha M \alpha^T}
= \text{ Pf } M
\end{equation}
and coincides with (\ref{Pf}).

\subsection*{Reduction to 3d}

Also in this case we can dimensionally reduce the theories to 3d.
When the $SP(2N)$ theory is reduced on the circle
there is an extra superpotential term,
$\eta Y$, as explained in 
\cite{Aharony:2013dha}. Moreover one can flow to a pure 3d
 theory by assigning to two fundamentals an opposite large real mass.
 
In the confined case we use again the prescription of \cite{Csaki:2014cwa}:
when considering the theory on the circle 
we keep the same field content and interactions of the 4d case.
This theory can be further reduced to a pure 3d theory.
This is done by assigning the real mass to the meson consistently with
the masses of the fundamental flavors.
In this case there are two massless components in the low energy spectrum:
one of them corresponds to the reduced meson $M_{red.}$
of the theory with $2(N+1)$ flavors while the second massless
field is the component $M_{2N+3}^{2N+4}$. The superpotential of the 3d theory is 
$W=M_{2N+3}^{2N+4}$ Pf $M_{red.} $.
The term $M_{2N+3}^{2N+4}$ has the same quantum numbers of the  
electric monopole $Y$ parametrizing the Coulomb branch of the electric theory 
with $2(N+1)$ flavors in the pure 3d case. By pursuing this identification
the superpotential
of the dual theory is 
\begin{equation}
\label{Wsympl}
W = Y \text{ Pf } M_{red.}
\end{equation}

\subsection*{Brane interpretation}

We can interpret the reduction of the 4d  confining $SP(2N)$
theory with $2(N+2)$ flavors and of 
its confined phase in terms of D branes.

We consider the system discussed above. Here we focus on the case
with the O4 planes, a similar discussion applies to the case with 
the O6 plane.
We compactify the direction $x_3$ and T-dualize along this direction.
After T-duality we have a stack of $2N$ D3 branes stretched between 
an NS and NS' brane.

The D6 branes become D5 branes and the D0 instanton becomes a D1 brane,
extended along  $x_3$.
When this brane wraps the compact direction $x_3$ it encounters two orientifold planes,
because after T-duality the $O4^+$ 
splits into the pair  $(O3^+,O3^-)$ \cite{Hanany:2000fq}.
The first plane is fixed at $x_3=0$ and the second one is at the \emph{mirror} point $\pi \alpha'/r$ \cite{Amariti:2015mva}.
Here we consider a situation with $2(N+2)$ D5 branes such that 
the dual theory, obtained by HW transition, becomes
an $SP(0)$ theory, i.e. there are no D3 left between the NS branes. 
We can study the contribution of the D1  branes to the theory or of their magnetic duals, corresponding to the
F1 strings. Their contribution to the effective action descends from the contribution of the 
stringy instanton, corresponding to the superpotential (\ref{WSP}).

The flow to the pure 3d limit follows from integrating out two fundamentals
with large real mass. 
On the brane side this real mass is obtained by moving two D5 branes 
at the mirror point
on the T-dual circle.
The electric theory at $x_3=0$
is an $SP(2N)$ theory with $2(N+1)$ fundamentals.
At the mirror point the D5s do not give any effect in the HW
transition because they cancel against the orientifold charge.
In this case there is an $\eta Y$ superpotential but no extra sectors.
In the dual picture moving the two D5s at the mirror
point has the same effect discussed in \cite{Amariti:2015mva}: 
this effect becomes a scale matching on the mesonic superpotential.
The extra D5 branes at the mirror point break  the meson into two massless components. 
This breaking induces the superpotential (\ref{Wsympl}).
This interaction involves the massless fields and it has to be considered also in the large mass limit.
\subsection*{The partition function}

Here we study the effects of the reduction of the confining $SP(2N)$ theory on the
superconformal index. The reduction of the index to the partition
functions for symplectic theories appeared in \cite{Aharony:2013kma,Gahramanov:gka}. 
In the case with $2(N+2)$ fundamentals 
the relation between the squashed three sphere partition functions of the two phases 
considered on the circle is
\begin{equation}
\int \prod_{i=1}^{N} 
d \sigma_i 
\left(\prod_{a=1}^{2(N+2)} 
\Gamma_h(\pm \sigma+\mu_a)
\!
\right)
\!
\Gamma_h^{-1}(\pm 2 \sigma_i) \!\!\!
\prod_{1\leq i<j \leq N}
\!\!\!\Gamma_h^{-1}(\pm \sigma_i \pm \sigma_j)
= \!\!\!\!\!\!\!
\prod_{1\leq a<b \leq 2(N+2)} 
\!\!\!\!
\Gamma_h (\mu_a + \mu_b)
\end{equation}
where we have to enforce the constraint $\sum \mu_a = 2 \omega$.
This constraint corresponds to the presence of the
 superpotential $\eta Y$ in the electric theory and to the 
 superpotential (\ref{WSP}) in the dual phase.
The  flow that reduces this duality to a pure 3d one 
is obtained by assigning the real masses as 
\begin{equation}
\mu_a =
\left\{
\begin{array}{lr}
~~m_a + m_A + \omega \Delta_1 &  a =1,\dots, 2N+2 \\
~~m ~- (N+1) m_A + \omega (1-(N+1) \Delta_1) \\
-m ~- (N+1) m_A + \omega (1-(N+1) \Delta_1) \\
\end{array}
\right.
\end{equation}
and computing the large $m$ limit.
The expected partition function for the $SP(2N)$ theory with $2(N+1)$ fundamental flavors and without the 
superpotential $\eta Y$ is obtained in the large $m$ limit.
This is computed by using formula (\ref{kargeZmass}). 
In the magnetic theory we obtain the contribution of the reduced meson $M_{red.}$ and, in addition, we have an extra term of the form 
$$
\Gamma_h (
2 \omega(1-(N+1)\Delta_1) - 2 (N+1) m_A)
= \Gamma_h(2 \omega - \sum \mu_a)
$$
where $\sum \mu_a = 2 (N+1) m_A$.
This corresponds to the contribution of the electric monopole $Y$ acting as a singlet in the dual phase.
\section{Conclusions}
\label{sec:conc}

In this paper we studied the reduction to 3d of a class of 4d s-confining theories, 
in the field theory and in the string theory regime, and we obtained 3d dualities.
In the string theory regime the structure of the 4d interaction of the confined phase is determined by an exotic D instanton configuration:  this contribution  corresponds to  the effective T-dual contribution of a D1 brane, when the 
compactification circle is kept at finite size.
We checked also the validity of the dualities  by studying the reduction of the  4d superconformal index to the 3d partition function on the squashed three sphere.

In this paper we did not discuss the reduction of orthogonal theories. 
It would be interesting to perform the analysis in the  $SO(N)$ case for both even and odd $N$. 
Another interesting aspect regards the reduction of the $\mathcal{N}=2$ stringy instanton studied in \cite{
Ghorbani:2010ks,Ghorbani:2011xh,Argurio:2012iw,Ghorbani:2013xga}.

\section*{Acknowledgments}
We are grateful to Claudius Klare and Alberto Mariotti for comments on the draft.
 A.~A.~is funded by the European Research Council (\textsc{erc}-2012-\textsc{adg}\_20120216) and acknowledges support by \textsc{anr} grant 13-\textsc{bs}05-0001. 
 A.~A.~would like to thank \textsc{ccny}, \textsc{ucsd}, Milano-Bicocca and Bern University for hospitality during various stages of this work.

\bibliographystyle{JHEP}
\bibliography{BibFile}

\end{document}